\documentclass[aps,reprint,showpacs,superscriptaddress,
footinbib,citeautoscript]{revtex4-1}

\usepackage{graphicx}

\usepackage[utf8]{inputenc}
\usepackage{csquotes}
\usepackage[american]{babel}
\usepackage[T1]{fontenc}
\usepackage{enumerate}
\usepackage{mdwlist}
\usepackage[activate=normal]{pdfcprot}
\usepackage{bbding}
\usepackage{color}
\usepackage{amssymb}
\usepackage{amsmath}
\usepackage{amsfonts}
\usepackage{mathrsfs}

\usepackage{bm}
\usepackage{dcolumn}
\usepackage{color}
\usepackage[colorlinks=true,citecolor=blue]{hyperref}
\hypersetup{colorlinks=true,citecolor=blue,linkcolor=red
,urlcolor=blue}

\begin{document}

\title{Plasmon modes of bilayer molybdenum disulfide: A density functional study}

\author{Zahra Torbatian}
\affiliation{School of Nano Science, Institute for Research in Fundamental Sciences (IPM), Tehran 19395-5531, Iran}
\author{Reza Asgari}
\affiliation{School of Nano Science, Institute for Research in Fundamental Sciences (IPM), Tehran 19395-5531, Iran}
\affiliation{School of Physics, Institute for Research in Fundamental Sciences (IPM), Tehran 19395-5531, Iran}

\begin{abstract}
We explore the collective electronic excitations of bilayer molybdenum disulfide (MoS$_2$) using the density functional theory together with the random phase approximation. The many-body dielectric function and electron energy-loss spectra are calculated
using an {\it ab initio} based model involving material-realistic physical properties. The electron energy-loss function of bilayer MoS$_2$ system is found to be sensitive to either electron or hole doping and it is owing to the fact that the Kohn-Sham band dispersions are not symmetric for energies above and below the zero Fermi level. Three plasmon modes are predicted. A damped high-energy mode, one optical mode (in-phase mode) for which the plasmon dispersion exhibits $\sqrt q $ in the long wavelength limit originating from low-energy electron scattering and finally a highly damped acoustic mode (out-of-phase mode).
\end{abstract}

\pacs{73.20.Mf, 71.10.Ca, 71.15.-m, 78.67.Wj}
\maketitle

\section{Introduction}\label{sec:intro}
Despite being the most promising two-dimensional (2D) material, gapless graphene has limitations of its applications in nanoelectronics and nanophotonics. This leads to finding of other 2D-materials with finite band gap such as transition metal dichalcogenides systems (TMDCs)~\cite{wang,TMD-DF2}.
Belonging to the family of layered TMCD, molybdenum disulfide (MoS$_2$) has been widely used in
 numerous areas, such as hydrodesulfurization catalyst, photovoltaic cell, photocatalyst, nanotribology, lithium battery,
 and dry lubrication, due to their distinctive electronic, optical, and catalytic properties~\cite{mos2}.
 In monolayer MoS$_2$, a strong photoluminescence peak at about $1.90$ eV, together with peaks at about $1.90$ and $2.05$ eV
 of the adsorption spectrum, indicates that MoS$_2$ undergoes an indirect to direct band gap transition when its bulk or
 multilayers is replaced by a monolayer~\cite{DFT-GW}.
Therefore, MoS$_2$ becomes a very interesting material owing to unique electronic and optical properties~\cite{vasp,ref_4-vasp,ref_5-vasp,ref_6-vasp}.

The collective density oscillations of doped two-dimensional crystalline systems have
recently received attention~\cite{plasmon, asgari, wang2}. Interestingly, the propagation of graphene plasmons has been directly imaged in real space by using scattering-type scanning near-field optical microscopy~\cite{fei} in which the wavelength of the plasmon is much smaller than the free-space excitation wavelength, allowing an extreme concentration of electromagnetic energy. This allows that graphene plasmon properties can easily be gate tuned.

Moreover, the energy of the metallic plasmon mode is restricted by the achievable carrier concentration and this limits the application of graphene plasmonics to the tetrahertz regime. Metallic TMDCs have much higher charge carrier densities stemming from a quite flat band structure leading to plasmon energies of around 1 eV in bulk TMDCs \cite{ref_14-An,ref_15-An}. Control and manipulation of the plasmon properties of these two-dimensional systems are expected to guide the design of next-generation nanophotonic and nanoelectronic devices, with the enhanced capability to operate from terahertz (THz) to infrared frequencies \cite{Si}.
\par
Recently the angle-resolved optical absorption and transmission in MoS$_2$ are studied by using current-current response tensor within the random phase approximation (RPA) \cite{Despoja}. Also, the optical properties of MoS$_2$ which include quasiparticle GW corrections, solving the Bethe-Salpeter equation, have been calculated  \cite{GW1,GW2}. Their results are in very good agreement with those measured data in the experiment.
The plasmon dispersions of monolayer MoS$_2$ and other TMDCs have been already calculated by Scholz et al., \cite{Scholz} using a model Hamiltonian and Andersen et al., \cite{Andersen} within {\it ab initio} simulations. While most of the aforementioned theoretical works have been focused on monolayer MoS$_2$, the plamon dispersion of bilayer MoS$_2$ has not been investigated yet.

In this paper, we have been using a recently proposed theoretical formulation \cite{MLG,BLG} based on {\it ab initio} density functional theory (DFT) together with the RPA to investigate the electronic excitation spectrum of bilayer MoS$_2$. For this purpose, the electronic ground-state of a periodically repeated slab is first determined and then a Dyson-like equation is solved within the RPA to calculate the density-density response function. Furthermore, a two-dimensional correction is applied to eliminate the artificial interaction between the replicas.
\par
The dynamical density-density response function of monolayer and bilayer MoS$_2$ are calculated within the DFT and RPA theory. Having known the density-density response function, we therefore can calculate the macroscopic dielectric function whose imaginary part gives the optical absorption spectrum and the collective modes are established by the zero in the real part of the macroscopic dielectric function.
Besides, not only the dielectric function is relevant for plasmon modes, but also for transport and the phonon spectra \cite{phonon} is a useful quantity. In addition, the dielectric function also provides the connection between theory and experiment. The theoretical dielectric function is related to the electron energy-loss function, and it provides useful information about the optical properties of the system. Here, we are just interested in the low-energy excitation investigating the collective modes of the intraband excitations.

This paper is organized as follows.
In Sec.~\ref{sec:model}, we present the methodology used to calculate the ground-state of the system and also describe a full {\it ab initio} approach to calculate the plasmon modes, properties of two-dimensional systems, based on density functional theory, within the random phase approximation. In Sec.~\ref{sec:result} we present and describe collective mode results of bilayer MoS$_2$. Finally, we conclude and summarize our main results in Sec.~\ref{sec:concl}

\section{THEORY AND COMPUTATIONAL METHODS}\label{sec:model}
\subsection{DFT calculations}
\begin{figure}
\includegraphics*[width=8.4cm]{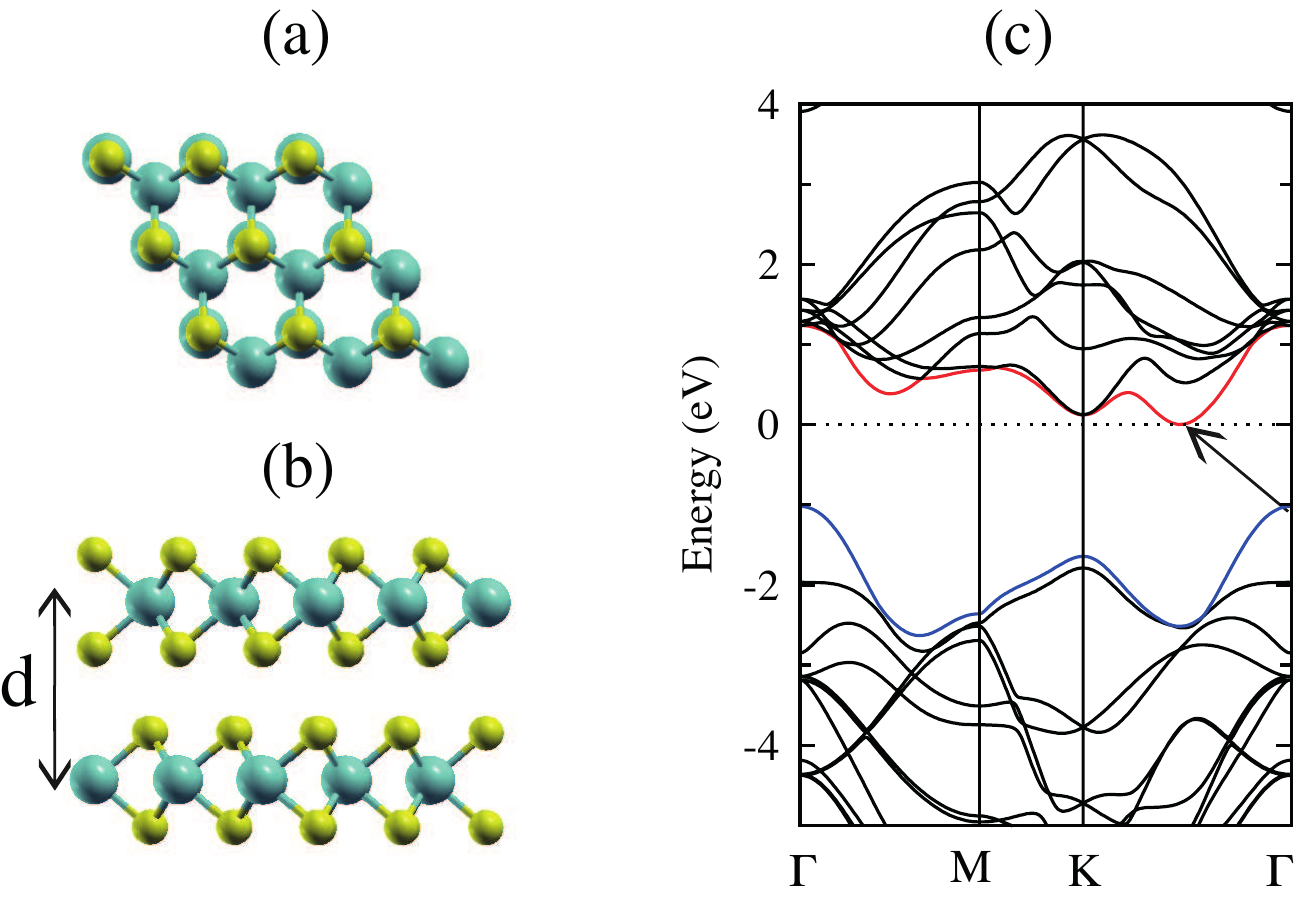}
\caption{(Color online) (a) Side view and (b) the first BZ of $AA'$ stacking structures of bilayer MoS$_2$ along the high symmetry $\Gamma-M-K-\Gamma$ directions. The definition of the layer distance $d$ is indicated. (c) Band structure of bilayer MoS$_2$.  The Fermi level is set at $0$ eV and $d=5.89$\AA. The arrow indicates the smallest value of the indirect band gap. The light blue balls are Mo atoms and the yellow ones are S atoms. The Mo-S bond and S-S band lengths are $2.39$ and $3.11$\AA, respectively}\label{fig1}
\end{figure}

In order to calculate the electronic structure, we use plane-wave basis in the local density approximation (LDA) and norm conserving pseudopotential within the Quantum Espresso (QE) package\cite{QE}.
The ground-state calculations are carried on unshifted $60\times60\times1$ Monkhorst-Pack (MP) K-point mesh of the first Brillouin zone (BZ) and with a $50$ Ry plane-wave cutoff. In the case of a layered material, we implement the van der Waals interaction. The nearest-neighbor in-plane and inter-plane distances are optimized, respectively, to the values $a=3.15$ and $d=5.89$\AA as defined in Fig. \ref{fig1}, however, for the unit cell constant, we use the experimental value of $a=3.16$\AA~\cite{aexp}. To avoid the effects of the interaction between images in the periodic unit cell calculations, the lattice parameter in the direction perpendicular to one single bilayer plane is chosen to be $L= 20$\AA. We find the $AA'$ stacking with Mo over S, as the most stable stacking order of bilayer MoS$_2$, which is the most studied in the previous calculation \cite{AA'}.

\par

\subsection{Density-density response function}

A central quantity in the theoretical formulation of the
many-body effects in electron liquid is the noninteracting
dynamical response function. The noninteractiong density-density response function of a three-dimensional periodic of electrons in the reciprocal space is given by
\begin{eqnarray}
\chi_{\mathbf G \mathbf G'}^{0}(q,\omega)=\frac{2}{\Omega}\sum_{\mathbf k,n,m}\frac{f_n(\mathbf k)-f_m(\mathbf k+\mathbf q)}{\hbar\omega+i\eta+\varepsilon_n(\mathbf k)-\varepsilon_m(\mathbf k+\mathbf q)}\nonumber\\
M_{n\mathbf k,m\mathbf k+\mathbf q}(\mathbf G)M^*_{n\mathbf k,m\mathbf k+\mathbf q}(\mathbf G')
\label{response}
\end{eqnarray}
which is a consequence of the Kubo formula for periodic the system \cite{kubo}.

Here, the factor of 2 accounts for the spin degeneracy and $f_n(\mathbf k)=\theta(\varepsilon_{\rm F}-\varepsilon_n(\mathbf k))$ is the Fermi-Dirac distribution of the charge carrier with energy $\varepsilon_n(k)$ at $T=0$. In the theory, the linear combination of plan-waves are used to determine the Kohn-Sham (KS) single-particle orbitals of the DFT. The KS wavefunctions are normalized to unity in the crystal volume $\Omega$. The broadening parameter $\eta$ used in this calculation is $0.02$ eV. In the summation over k, we use $151\times151\times1$ K-point mesh sampling in the BZ and the $n$ and $m$ sum over $20$ bands for monolayer and $30$ bands for bilayer MoS$_2$.
\par
The density-density response function can be obtained in the framework of DFT, as follows \cite{DFT}:
\begin{eqnarray}
\chi_{\mathbf G\mathbf G'}=\chi^0_{\mathbf G\mathbf G'}+\sum_{{\bf G_1 G_2}}\chi^0_{\mathbf G\mathbf G_1}\nu_ {\mathbf G_1\mathbf G_2}\chi_{\mathbf G_2\mathbf G'}
\end{eqnarray}
where $\nu_{\mathbf G\mathbf G'}$ represent the Fourier coefficients of an effective electron-electron interaction. For the electron liquid the bare Coulomb interaction is given by $\nu_{\mathbf G\mathbf G'}^0=4\pi e^2\delta_{\mathbf G\mathbf G'}/|\mathbf q+\mathbf G|^2$. The RPA procedure, an approximation valid in the high-density limit, takes into account electron interaction only to
the extent required to produce the screening field and thus the response to
the screened field is measured by $\chi^0$.

The matrix elements of Eq. (\ref{response}) have the form
\begin{eqnarray}
M_{n\mathbf k,m\mathbf k+\mathbf q}({\mathbf G})=<\Phi_{n\mathbf k}|e^{-i(\mathbf q+\mathbf G)\cdot \mathbf r}|\Phi_{m\mathbf k+\mathbf q}>_{\Omega}
\end{eqnarray}
where $\mathbf q$ is the momentum transfer vector parallel to the $x-y$ plane. Wave functions $\Phi_{n\mathbf k}(\mathbf r)$ are the KS electron wave functions expanded in the plane-wave basis have the form
\begin{eqnarray}
\Phi_{n\mathbf k}(\mathbf r)=\frac{1}{\Omega}\sum_{\mathbf G}C_{n\mathbf k}(\mathbf G)e^{i(\mathbf k+\mathbf G)\cdot\mathbf r}
\end{eqnarray}
where the coefficients $C_{n\mathbf k}(\mathbf G)$ are obtained by solving the LDA-KS equations self-consistently.

This approach is correct for the purely three-dimensional periodic system. The long-range behavior of the Coulomb interaction allows non-negligible interactions between repeated planar arrays even at large distance. This unphysical phenomenon can be removed by replacing $\nu_{\mathbf G\mathbf G'}$ by the truncated Fourier
integral over the cut of plane axis $(z)$ \cite{MLG,BLG,NRG} and thus we have
\begin{eqnarray}
\nu^0_{\mathbf G\mathbf G'}=\frac{2\pi e^2\delta_{\mathbf g\mathbf g'}}{|\mathbf q+\mathbf g|}\int_{-L/2}^{L/2}dz\int_{-L/2}^{L/2}dz'e^{i(G_{z'}z'-G_zz)-|\mathbf q+\mathbf g||z-z'|}\nonumber\\
\end{eqnarray}
where the $\mathbf g$ and $G_z$ denote the in-plane and out-plane components of $\mathbf G$ and we assume that $q$ is never be zero owing to a uniform background of positive charge.

In the framework of linear response theory, the inelastic cross section corresponding to a process where the external perturbation creates an excitation of energy $\hbar \omega$ and wavevector $\mathbf q+\mathbf G$ is related to the diagonal elements of the dielectric function in the level of the RPA
\begin{eqnarray}
\epsilon_{\mathbf G\mathbf G'}=\delta_{\mathbf G \mathbf G'}-\sum_{\mathbf G_1} \nu^0_{\mathbf G\mathbf G_1}\chi^0_{\mathbf G_1 \mathbf G'}
\end{eqnarray}
and the plasmon modes are established by the zero in the real part of the macroscopic dielectric function given by
\begin{eqnarray}
\epsilon(q,\omega)=\frac{1}{(\epsilon^{-1})_{\mathbf G\mathbf G'}}|_{\mathbf G=\mathbf G'=0}
\end{eqnarray}
as long as there is no damping.

Electron energy-loss spectroscopy (EELS), on the other hand, is an analytical technique which is based on inelastic scattering of fast electrons in a thin sample. EELS offers unique possibilities for advanced materials analysis owing to the broad range of inelastic interactions of the high energy electrons with the specimen atoms, ranging from phonon interactions to ionisation processes. The low-loss or valence region of an electron energy-loss function (EEL) spectrum provides similar information to that provided by optical spectroscopy, containing valuable information about the band structure and in particular about the dielectric properties of a material e.g., band gap, surface plasmon modes. Interestingly, the most prominent peak comes from a plasma resonance of the valence atoms. The EEL is proportional to the imaginary part of the inverse dielectric function which is given by
\begin{eqnarray}
E_{EEL}(q,\omega)=-\Im m[1/\epsilon(q,\omega)]
\end{eqnarray}
It is worth mentioning that the nonlocal field effects are included in EEL through the off-diagonal elements of the general $\chi_{\mathbf G \mathbf G'}$\cite{nlc} function.
\par
\section{RESULTS AND DISCUSSION}\label{sec:result}

In this section, we present our main numerical results based on first-principles simulations. Our aim is to explore the density-density response function and plasmon modes of bilayer MoS$_2$. All the first-principles calculations are performed at zero temperature without considering the spin-orbit interaction. Our study is devoted to a consideration of the homogenous electron liquid
in both the high-density and the intermediate-density regime. We
do not consider the low-density charge carrier, primarily because its
behavior is not especially relevant to the properties of metals.

To begin with, we examine the electronic structure and plasmon modes of monolayer MoS$_2$ within the aforementioned theory. Our numerical results of the plasmon modes are in very good agreement with those results obtained in \cite{welling}. It would be worth mentioning that the plasmon mode obtained within DFT-RPA approach for monolayer MoS$_2$ differs with that calculated using the low-energy model Hamiltonian in \cite{Scholz} and reasons may explain by having the multi-orbital and
multiband structures of the system which discussed in~\cite{asgari2, louie}, thought the plasmon modes calculated by DFT-RPA approach for monolayer and bilayer graphene are in good agreement with those results obtained using the low-energy model Hamiltonian. Thus it would pretty accurate to calculate the collective modes based on the DFT-RPA specially for complex systems which contain several bands and orbitals. Having investigated these examinations, we can now survey bilayer MoS$_2$ and explore its ground-state electronic properties and charge collective excitations.

\subsection{Electronic structures of bilayer MoS$_2$}
The band structure of bilayer MoS$_2$ along the high symmetry $\Gamma-M-K-\Gamma$ directions of the BZ is shown in Fig. \ref{fig1}. The blue and red lines correspond to the valence and conduction bands, respectively.
\par
In contrast to monolayer MoS$_2$, bilayer MoS$_2$ is an indirect gap semiconductor that the maximum of the valence band is located at the $\Gamma$ point of the BZ while the conduction band minima is found between the $\Gamma$ and $K$ high symmetry points (Q point) \cite{BL}. The conduction band states at K point are mainly owing to the localized $d$ orbitals of the Mo atoms, located in the middle of the S-Mo-S layer sandwiches and relatively unaffected by interlayer coupling. However, the states near the $\Gamma$ point are due to the combinations of the antibonding $p_z$ orbitals of S atoms and the $d$ orbitals of Mo atoms and have a strong interlayer coupling effect. Therefore, as the layer number changes, the direct excitonic states near the K point are relatively unchanged. The transition at the $\Gamma$ point shifts significantly from indirect gap to a larger and direct one. Therefore, changing in the band structure of layer number is due to quantum confinement and the resulting change in hybridization between the $p_z$ orbitals of S atoms and $d$ orbitals of Mo atoms \cite{kadan}.
\par
\begin{figure}
\includegraphics[width=8.4cm]{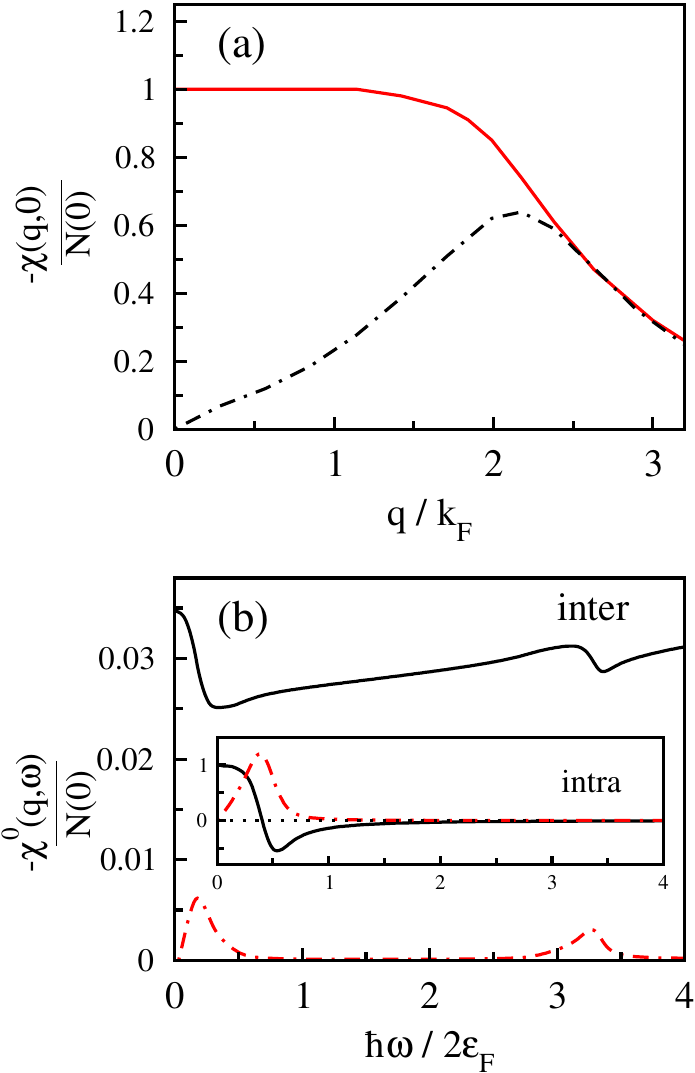}
\caption{(Color online) (a) The static response function in units of the Fermi-level density of states as a function of $q/k_{\rm F}$. Noninteracting and interacting density-density response function represent by solid and dash-dotted lines, respectively. (b) The intra and inter contributions of the noninteracting response function as function of $\hbar \omega/2\varepsilon_{\rm F}$. The real part (black) and the imaginary part (red) of the dimensionless function $\frac{-\chi^0(\mathbf q, \omega)}{N(0)}$ for $q = 0.6 k_{\rm F}$. Solid line represents $-\Re e\chi^0(q,\omega)/N(0)$ and dashed-dot line refers to $-\Im m\chi^0(q,\omega)/N(0)$. Notice that the peak of the imaginary part is corresponding to a zero in the real part of the density response function.
}\label{fig2}
\end{figure}

\subsection{Collective excitation spectra in extrinsic bilayer MoS$_2$}
We investigate the dielectric properties of doped bilayer MoS$_2$. The extrinsic systems are simulated by slightly changing the level populations in Eq. (\ref{response}) and assuming that the doping has a negligible effect on the KS electronic structure.

The static limit of the response function $\chi^0 (\mathbf q, \omega=0)$, which contains a number of noteworthy features, is purely real as illustrated in Fig. \ref{fig2}(a) in units of the noninteracting density of states at the Fermi surface for electron doping with $E_{\rm F}=0.06$ eV. The $\lim_{q \to 0} \chi^0(\mathbf q, \omega=0$) is finite and equal to the density of states at the Fermi energy, $N(0)$ a measure of the number of excited states. Apparently, the structure of the static noninteracting response function, in the vicinity of the $q=2k_{\rm F}$, is very similar to that found in three-dimensional electron gas. The static limit of the response function exhibits a Kohn anomaly at $q=2k_{\rm F}$ and it is responsible for several interesting phenomena such as Friedel oscillations. Basically, the static response function $\chi^0 (\mathbf q, \omega=0)$ can be Fourier transformed to obtain $\chi^0 (\mathbf r, \omega=0)$, which physically represents the response of the density of the noninteracting electron gas to a potential that is located at the origin of the coordinate system \cite{vignale}. Furthermore, we plot the density-density response function $\chi(\mathbf q, \omega=0)=\chi^0 (\mathbf q, 0)/\epsilon(q,0)$ in the same figure. The interacting response function is different from the noninteracting one. The difference with the $\chi^0 (\mathbf q, \omega=0)$ is noticeable in the long wavelength limit, where the diverging dielectric function makes $\chi(\mathbf q, \omega=0)$ vanish as $q$. Also, the real part (black) and the imaginary part (red) of the noninteracting response function $-\chi^0(\mathbf q, \omega)/N(0)$ as a function of $\hbar \omega$ are plotted in Fig. \ref{fig2}(b) for $q = 0.6 k_{\rm F}$. As we expected the peak of the imaginary part of the response function, $\Im m\chi^0(\mathbf q, \omega)$ is corresponding to a zero in the real part of the response function, $\Re e\chi^0(\mathbf q, \omega)$. Notice that $\Re e\chi^0(q,\omega)$ changes sign from negative to positive as $\omega$ moves across the electron-hole continuum. In addition, the density-density response function originates from the inter- and
intra-band contributions to this quantity. There are three different peaks in those quantities. One strong peak of the intra-band contribution refers to a main collective mode and two others which are small and may lead two different plasmon modes at small and high energy regions. Basically, the effect of those peaks would have appeared in the electron loss function and we will discuss this point later.

\begin{figure}
\includegraphics[width=7.7cm]{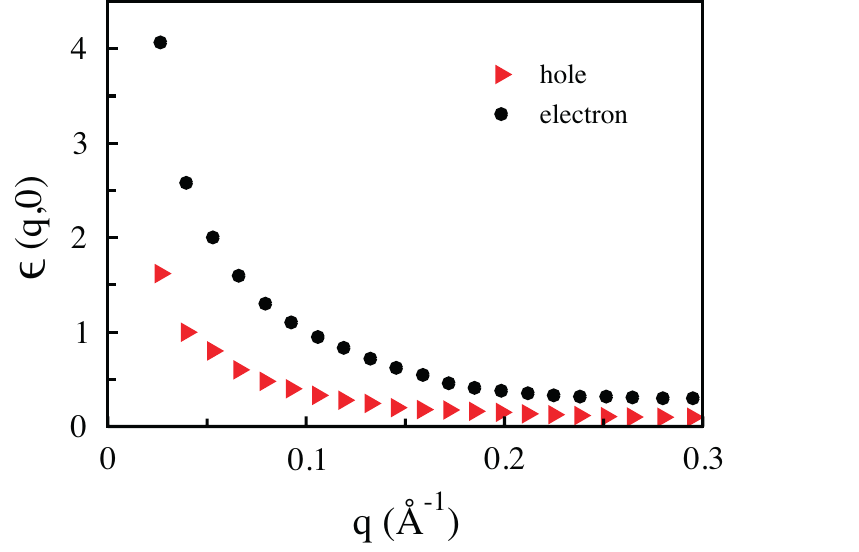}
\caption{(Color online) Static dielectric function of bilayer MoS$_2$ along the high-symmetry $\Gamma K$ direction for electron and hole doping. The value of concentration is $n_e=n_h= 6.9 \times 10^{13}$ cm$^{-2}$ for the both cases. The many-body screening of the electron doping is greater than that of the hole doping case.}\label{fig3}
\end{figure}

In Fig. \ref{fig3}, we plot the static dielectric function, $\epsilon(\mathbf q, 0)$ of bilayer MoS$_2$ for electron and hole doped cases where the charge density is $6.9 \times 10^{13}$ cm$^{-2}$. Interestingly enough, the $\epsilon(\mathbf q, 0)$ for the hole doped case is smaller than the electron doped one and shows a stronger many-body screening in the electron doped case. There is a clear dependence of the static dielectric function with the type of doping, showing a different behavior for electron or hole doping.

Here, we study the collective modes of bilayer MoS$_2$, which are defined by the zeroes of the dielectric function $\epsilon(\mathbf q, \omega)$. The dispersion relation of the plasmon modes is defined from Re $\epsilon (\mathbf q, \omega)=0$ which leads to poles in the EEL function $-\Im m\frac{1}{\epsilon(\mathbf q, \omega)}$, that can be measured by means of  the electron-loss spectroscopy.
\par
\begin{figure}
\includegraphics[width=7.7cm]{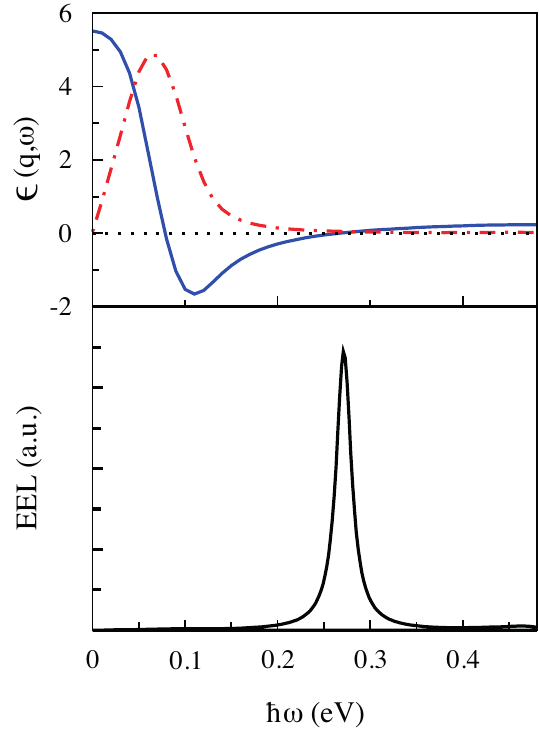}
\caption{(Color online) (upper panel) Calculated real and imaginary parts of the macroscopic dielectric function of bilayer MoS$_2$, refer to solid and dashed line respectively, for momentum transfer $0.079$\AA$^{-1}$ and concentration of $n_e= 6.9\times$10$^{13}$ cm$^{-2}$ along $\Gamma K$ direction. (bottom panel) EEL as a function of the energy $\hbar \omega$. The peak of the EEL corresponding to a zero in the many-body dialectic function. }\label{fig4}
\end{figure}

We begin with Fig. \ref{fig4} (upper panel), which exhibits the real and imaginary parts of the dielectric function together with their electron energy-loss function of bilayer MoS$_2$  and momentum transfer along $\Gamma K$ direction, namely $q=0.079$\AA$^{-1}$ for $E_{\rm F}= 0.05$ eV ($n_e= 6.9 \times10^{13}$cm$^{-2}$). Notice that value of the $\Re e\epsilon(q,\omega)$ or $\Im m\epsilon(q,\omega)$ depends strongly on the Fermi energy. Moreover, there is a value of $\omega$ for which both $\Re e\epsilon(q,\omega)$ and $\Im m\epsilon(q,\omega)$ are zero, which it corresponds to the plasmon mode with a long lived mode of the system. Furthermore, as shown in the Fig. \ref{fig4} numerically (bottom panel), the peak in the electron energy-loss spectra corresponding to a zero in the real part of the dielectric function, at a frequency where the imaginary part, $\Im m\epsilon(q,\omega)$ is almost zero.

\begin{figure}
\includegraphics[width=7.7cm]{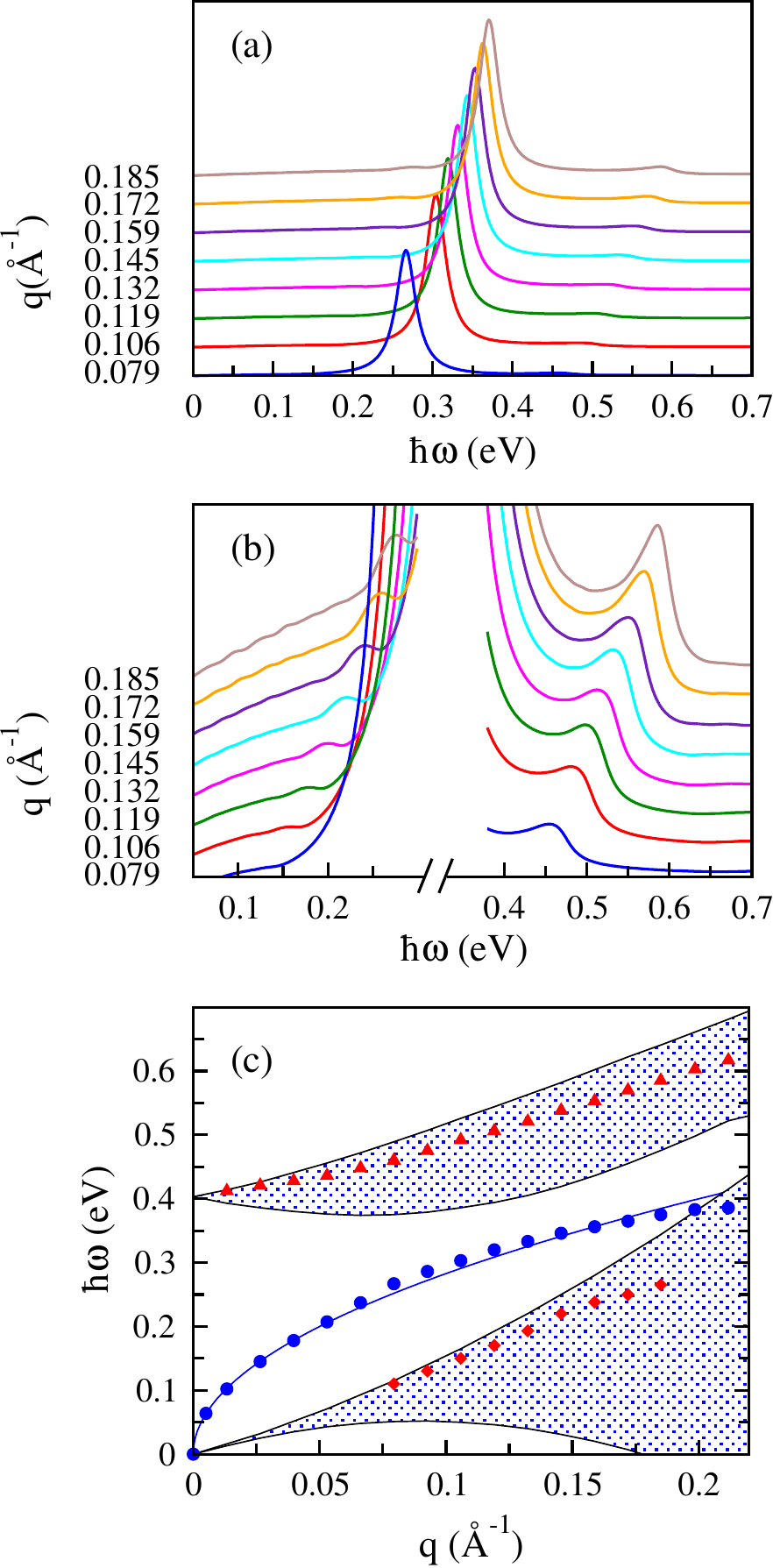}
\caption{(Color online) EEL function of bilayer MoS$_2$ for different momentum transfers $\mathbf q$ for the electron concentration of $n_e=6.9\times 10^{13}$cm$^{-2}$ along the high-symmetry $\Gamma K$ direction. (b) The same as (a) for two different energy regions. (c) The three different plasmon dispersions obtained from the peaks in the EEL functions in the (a) and (b) figures. The solid line shows a $\sqrt{q}$ fitted function for the optical plasmon mode. The blue dots refer to the optical plasmon mode, the red triangle and diamond symbols refer to the high energy and acoustic plasmon modes, respectively. The shaded areas represent the electron-hole
continuum}\label{fig5}
\end{figure}

In Fig. \ref{fig5}(a), we display the EEL of bilayer MoS$_2$ for the electron doping value $n_e= 6.9 \times 10^{13}$ cm$^{-2}$ and for energies $\hbar \omega$ below 1 eV and momenta $\mathbf q$ along the high-symmetry $\Gamma K$ direction. It is clear that the EEL is dominated by a narrow peak, which is placed at $\omega \thicksim 0.25-0.35$ eV. Furthermore, we explore other peaks in the system by studying the EEL at small and large energies as shown in Fig. \ref{fig5}(b). There are two peaks in the EEL, which originate from the inter-band contribution of the response function, leads to extra collective modes. Moreover, the $\Im m\epsilon(q,\omega)$ at position of those peaks are finite. Therefore, it turns out that those collective modes are damped and the damping parameter, $\gamma$ would be momentum dependent. In this case, we should verify the condition where $\epsilon(q,\omega_p-i\gamma)=0$ is satisfied. This equations leads to two separate equations in which the $\gamma$ is defined by $\gamma(q)=\Im m\chi^0(q,\omega)/\partial\Re e\chi^0(q,\omega)/\partial\omega$ at $\omega=\omega_p$ and the collective mode is obtained by $1-\nu_q\Re e\chi^0(q,\omega)-\gamma\nu_q\partial\Im m\chi^0(q,\omega)/\partial\omega=0$ again at $\omega=\omega_p$.

We plot three plasmon dispersions of the system obtained from the peaks in the EEL in Fig. \ref{fig5}(c). A damped high-energy mode, one optical mode (in phase) for which the plasmon dispersion exhibits $\sqrt q $ in the long wavelength limit originating from low-energy electron scattering and finally a highly damped acoustic mode (out-of-plane) are shown in this figure. The shaded areas show the electron-hole continuum, the region of the $q, \omega$ plane in which $\Im m \chi^0(q,\omega)$ differs from zero, that its boundary can be obtained at any specific momentum transfer $\mathbf q$ by the difference between system energy at $\mathbf{k_{\rm F}}$ and $\mathbf{k_{\rm F}+q}$. Besides, two damped modes lie on the electron-hole continuum regions. In the optical mode and for longer momentum transfer, the strength of the plasmon is significantly reduced owing to the screening by interband transitions \cite{Scholz, Andersen}.
Notice that as one goes to larger values of $q$, the optical plasmon energy increases rather slowly; the maximum single-pair energy increases more rapidly. Furthermore, the optical plasmon oscillator strength vanishes as its dispersion approaches the edge of the electron-hole continuum at $q=q_c$. In a vicinity of $q\rightarrow q_c$, the quantity $|\partial \epsilon(q,\omega)/\partial \omega|$ tends to infinity too.

\begin{figure}
\includegraphics[width=7.7cm]{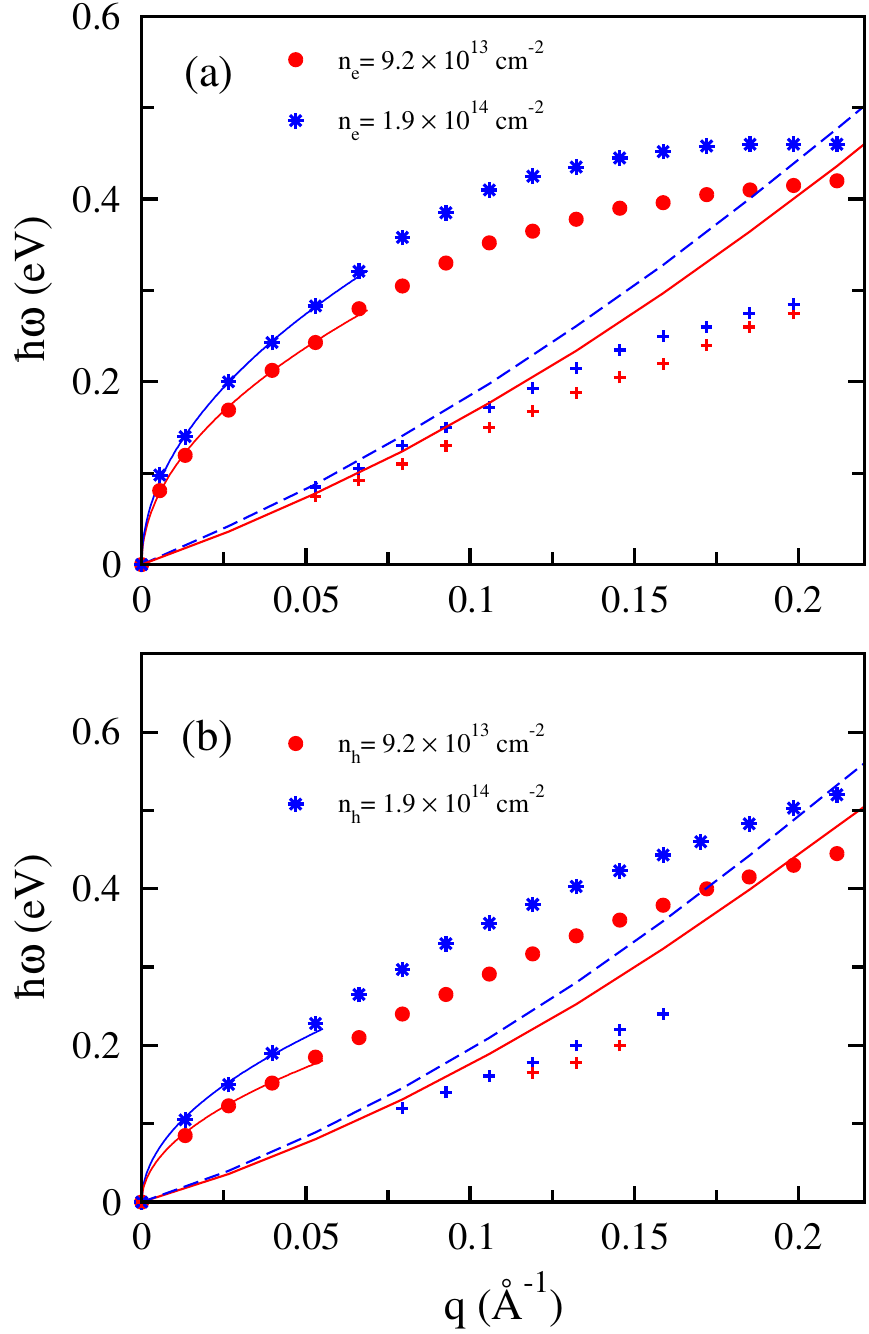}
\caption{(Color online) The plasmon dispersions of the electron (a) and hole (b) doped bilayer MoS$_2$ for two concentrations ($19$ and $9.2$ $\times$10$^{13}$cm$^{-2}$) along the $\Gamma$K. The change of the plasmon modes in the hole doped is stronger than that in the electron doped one interns of doping. A $\sqrt{q}$ function is fitted to the long wavelength region for each optical mode and shown by a short solid line. Acoustic modes are located inside the electron-hole continuum and their boundaries are shown by the solid and dashed lines. Notice that the optical plasmon modes increase by doping the system.}\label{fig6}
\end{figure}

In Fig. \ref{fig6} the optical mode and damped acoustic dispersions are shown at a given carrier concentration for (a) electron and (b) hole doping (we neglect to show the other high energy mode). The electron-hole continuum are shown by the solid and dashed lines. It is worth mentioning that the plasmon modes in electron and hole doping graphene, at a given carrier concentration, show the same dynamics. However, it is quite different in bilayer MoS$_2$ \cite{Scholz}. More importantly, the change of the plasmon modes in the hole doped case is much stronger than that in the electron doped one at given density.
As Fig. \ref{fig1} shows, the structure of the valence bands is quite different compared to the conduction bands. Therefore, there is no electron-hole symmetry in MoS$_2$, and intriguingly the plasmon dispersion of electron and hole doped cases are completely different. The electron and hole effective band masses of monolayer and bilayer MoS$_2$ given in Table~\ref{tab:table1} that coincide with the effective masses reported in Ref.~\cite{Yun}. The effective band masses of bilayer MoS$_2$ for an electron in the conduction band minima at the Q point and for a hole at the valence band maximum at the $\Gamma$ point are determined to be $m_e^* = 0.56$ and $m_h^* = 1.07$, respectively, in units of the electron bare mass, that thus confirms the anisotropy of the valence and conduction bands. Furthermore, the plasmon modes depend on the Fermi energy values and increases by increasing the Fermi energy.

\begin{figure}
\includegraphics[width=7.7cm]{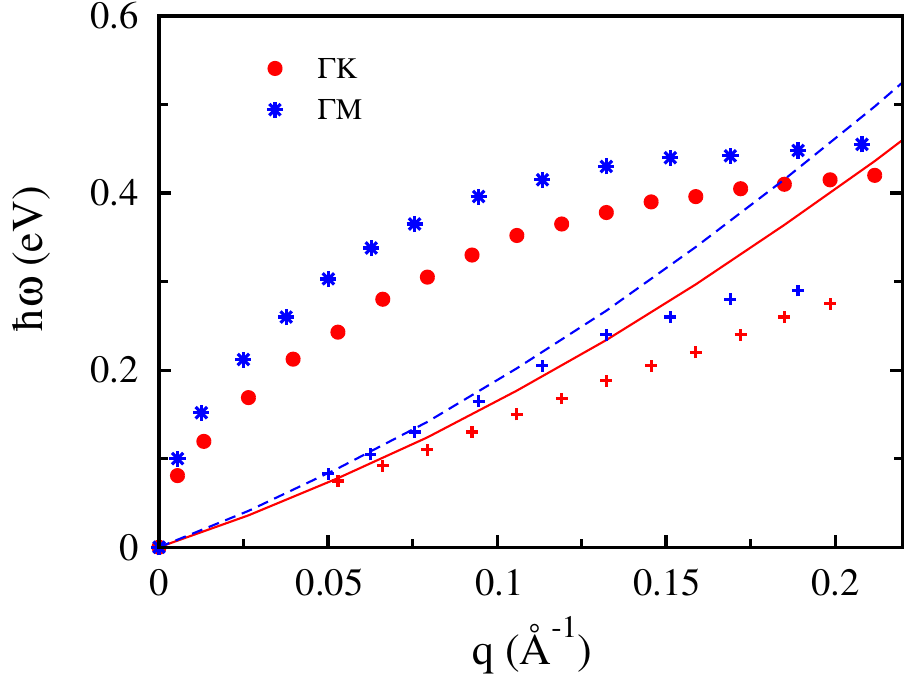}
\caption{(Color online) The plasmon dispersions of the electron doped bilayer MoS$_2$ for a momentum $\mathbf q$ along the $\Gamma K$ (blue) and $\Gamma M$ (black) directions for $E_{\rm F}= 0.067$ eV corresponding $n_e=9.2\times10^{13}$cm$^{-2}$. Acoustic modes are located inside the electron-hole continuum and their boundaries are shown by the solid and dashed lines.  }\label{fig7}
\end{figure}

\begin{figure}
\includegraphics[width=5.0cm]{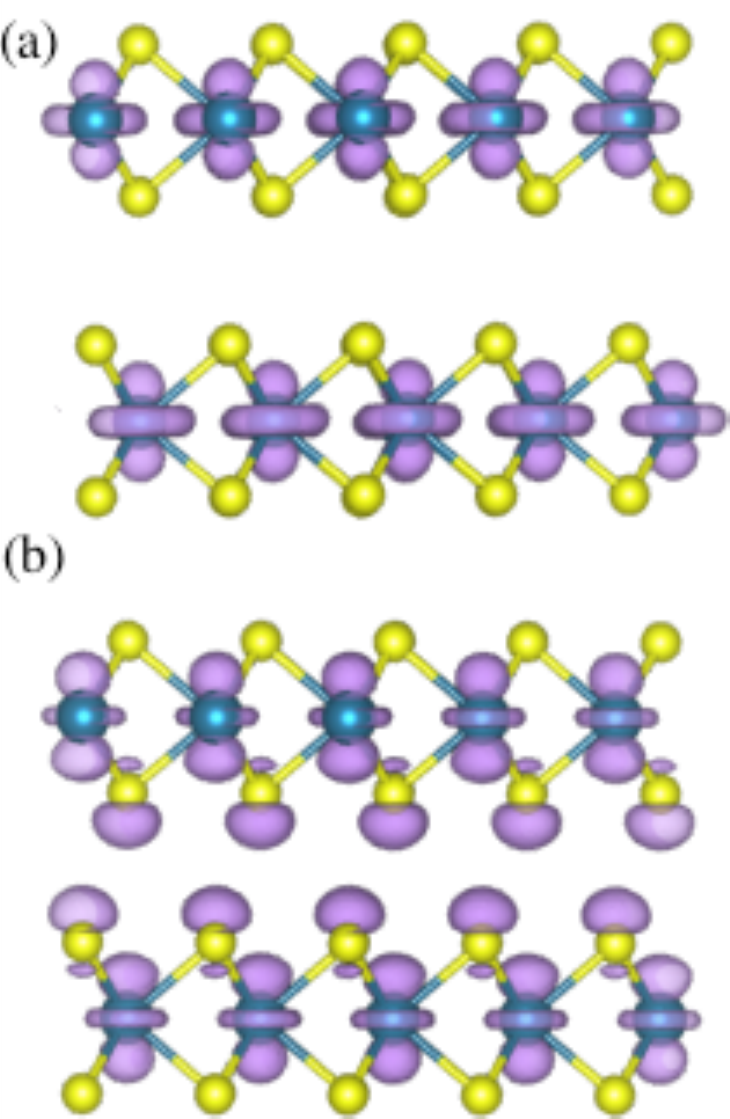}
\caption{Side view of electron charge density of (a) the CBM at K-point and (b)
 VBM states at $\Gamma$-point of bilayer MoS$_2$. The conduction band states at the $K$ point are in the Mo places, while the charge on the VBM accumulated on the inner S atoms. }\label{fig8}
\end{figure}

\begin{table}[!th]
\caption{\label{tab:table1}
Calculated the hole and electron effective masses ($m_h^*$, $m_e^*$) in units of the electron bare mass at the high symmetry points. The effective masses of MoS$_2$ are derived from the band structures as shown in Fig. \ref{fig1}}
\begin{ruledtabular}
\begin{tabular}{cccc}
& &\multicolumn{2}{c}{Effective Masses}\\ \cline{3-4}
Structure & point        & $m_e^*$  &  $m_h^*$      \\ \hline
Monolayer & K            &  0.47          &  0.61               \\
          & Q            &  0.58          &                     \\
          & $\Gamma$     &                &  3.5                \\
 Bilayer  & K            &  0.53          &  0.63               \\
          & Q            &  0.56          &                      \\
          & $\Gamma$     &                &  1.07                \\
\end{tabular}
\end{ruledtabular}
\end{table}

\par

For the sake of completeness, we compare effective band masses between monolayer and bilayer Mo$_2$ system. From monolayer to bilayer MoS$_2$, the $m_h^*$ at the $K$ point dramatically increases at the $\Gamma$ point. While $m_e^*$ at the $K$ point changes slightly at the Q point (Table~\ref{tab:table1}). The change of the conduction minimum band from the $K$ point to the $Q$ point may have no significant effect because of their similar effective masses. In contrast the change of the valence maximum band from the $K$ point in the $\Gamma$ point, induces the dramatic increase of the effective mass.

In order to explore the dependence of the collective modes on the symmetry direction in the BZ, we compare the plasmon dispersions (optical and acoustic modes) of electron-doped bilayer MoS$_2$ for $n_e=9.2\times10^{13}$ cm$^{-2}$ along the high-symmetry inequivalent directions $\Gamma M$ and $\Gamma K$ in Fig. \ref{fig7}. In particular, the plasmon dispersions along the $\Gamma M$ is energetically higher than the plasmon dispersions along the $\Gamma K$.

The  distribution of the charge densities of the valence band maximum (VBM) and conduction band minimum (CBM) states is shown in Fig. \ref{fig8} for bilayer MoS$_2$. The CBM states at the $K$ point are distributed within the Mo sublattice while the charge on the VBM at the $\Gamma$ point accumulated on the inner S atoms. This indicates that the states at the $\Gamma$ point have strong interlayer coupling, while the states around the $K$ valley remain largely unaffected. This may also explain why there is quantum effect difference in the plasmon mode for the electron and hole doped systems as shown in Fig. \ref{fig6}.

It is important to mention that when the extrinsic Fermi level crosses two bands, two Fermi surfaces per inequivalent K point, two collective modes in the spectrum emerge. Therefore, two different oscillations of charge carriers occur, leading to the appearance of two collective modes: one in which the oscillation is in phase (the optical mode) and another one is the out of phase oscillation (acoustic mode)~\cite{Eguiluz}. In our numerical study, different charge-carrier concentrations are considered such a way that the extrinsic Fermi levels cross only one band, for instance, there is only one intersection around $Q$ point in the electron doped case.

\section{CONCLUSION}\label{sec:concl}

We have used a full DFT simulations together with RPA analysis to calculate the energy loss functions and plasmon dispersions of extrinsic bilayer MoS$_2$ for energies up to 1 eV. The interacting density-density response function as well as dielectric function are calculated using the theory in $q, \omega$ plane. In extrinsic bilayer MoS$_2$, the electron energy-loss function is calculated and we have explored that the function is sensitive to doping is either positive or negative, as there is no electron-hole symmetry in MoS$_2$. Furthermore, we have calculated the collective modes of the system for various charge carrier densities. We have found three different collective modes in the system. One mode occur at high-energy and for which the mode is damped. Other mode refers to the optical mode in which the plasmon dispersion exhibits a $\sqrt q$ dispersion as the conventional plasmon of a two-dimensional electron gas originating from low-momentum carrier scattering. An acoustic mode in the system is observed however this mode is heavily damped. In addition, we have observed that the plasmon modes of electron and hole doping of bilayer MoS$_2$ are not equivalent and the discrepancy is owing to the fact that the Kohn-Sham band dispersions are not symmetric for energies above or below the zero Fermi level. Finally, we should emphasize that the material-specific dielectric function considering the multi-orbital and multiband structures are needed to obtain realistic plasmon dispersions in MoS$_2$. Our numerical finding can be investigated by current experiments.

\section{acknowledgments}
Z. T. would like to thank Iran Nanotechnology Initiative Council for their support. This work is also partially supported by Iran Science Elites Federation.

\end{document}